\documentclass{elsart}

\usepackage{graphicx}
\usepackage{amssymb}

\def\NIMA#1#2#3{Nucl. Instr. Meth. Phys. Res. A {\bf #1}\ (#2)\ #3}
\def\NIM#1#2#3{Nucl. Instr. Meth. {\bf #1}\ (#2)\ #3}
\def\IEEE#1#2#3{IEEE Trans. Nucl. Sci. vol. {\bf #1}\ (#2)\ #3}
\def\CPC#1#2#3{Comp. Phys. Comm. {\bf #1}\ (#2)\ #3}

\begin{document}

\begin{frontmatter}

\title{Electron/Pion Identification with ALICE TRD Prototypes
using a Neural Network Algorithm}

\oddsidemargin=2.5cm
\topmargin=2cm
\voffset -1.3in
\hoffset -1.in

\textwidth=16cm
\textheight=26cm
\sloppy

\linespread{1.}  

\vspace{.5cm}
\author[hei]{C.~Adler}, 
\author[gsi]{A.~Andronic}, 
\author[kip]{V.~Angelov},
\author[gsi]{H.~Appelsh{\"a}user}, 
\author[mue]{C.~Baumann}, 
\author[fra]{C.~Blume}, 
\author[gsi]{P.~Braun-Munzinger}, 
\author[mue]{D.~Bucher}, 
\author[gsi]{O.~Busch}, 
\author[buc]{V.~C{\u a}t{\u a}nescu}, 
\author[dub]{S.~Chernenko},
\author[buc]{M.~Ciobanu}, 
\author[gsi]{H.~Daues},
\author[hei]{D.~Emschermann}, 
\author[dub]{O.~Fateev},
\author[gsi]{Y.~Foka},  
\author[gsi]{C.~Garabatos}, 
\author[mue]{R.~Glasow},
\author[mue]{H.~Gottschlag},
\author[tok]{T.~Gunji}, 
\author[tok]{H.~Hamagaki}, 
\author[gsi]{J.~Hehner}, 
\author[mue]{N.~Heine},
\author[hei]{N.~Herrmann}, 
\author[tok]{M.~Inuzuka}, 
\author[dub]{E.~Kislov},
\author[hei]{T.~Lehmann}, 
\author[kip]{V.~Lindenstruth},
\author[gsi]{C.~Lippmann}, 
\author[hei]{W.~Ludolphs}, 
\author[hei]{T.~Mahmoud}, 
\author[gsi]{A.~Marin},  
\author[gsi]{D.~Miskowiec}, 
\author[hei]{K.~Oyama}, 
\author[dub]{Yu.~Panebratsev},
\author[hei]{V.~Petracek}, 
\author[buc]{M.~Petrovici},
\author[buc]{A.~Radu},
\author[mue]{K.~Reygers},
\author[hei]{I.~Rusanov}, 
\author[gsi]{A.~Sandoval},
\author[mue]{R.~Santo},
\author[hei]{R.~Schicker},
\author[gsi]{R.S.~Simon}, 
\author[dub]{L.~Smykov},
\author[hei]{H.K.~Soltveit}, 
\author[hei]{J.~Stachel}, 
\author[gsi]{H.~Stelzer}, 
\author[hei]{M.R.~Stockmeier},
\author[gsi]{G.~Tsiledakis}, 
\author[mue]{W.~Verhoeven},
\author[hei]{B.~Vulpescu}, 
\author[mue]{J.P.~Wessels}, 
\author[mue]{A.~Wilk\thanksref{info}}, 
\author[hei]{B.~Windelband},
\author[dub]{V.~Yurevich},
\author[dub]{Yu.~Zanevsky},
\author[mue]{O.~Zaudtke}

\vspace{2mm}
\address[hei]{Physikaliches Institut der Universit{\"a}t Heidelberg, Germany}
\address[gsi]{Gesellschaft f{\"u}r Schwerionenforschung, Darmstadt, Germany}
\address[kip]{Kirchhoff-Institut f\"ur Physik, Heidelberg, Germany}
\address[fra]{Institut f{\"u}r Kernphysik, Universit{\"a}t Frankfurt am Main, Germany}
\address[mue]{Institut f{\"u}r Kernphysik, Universit{\"a}t M{\"u}nster, Germany}
\address[buc]{NIPNE Bucharest, Romania}
\address[dub]{JINR Dubna, Russia}
\address[tok]{University of  Tokyo, Japan}

{\large for the ALICE Collaboration}

\thanks[info]{Corresponding author: Institut f{\"u}r Kernphysik, Wilhelm-Klemm-Str. 9, 48149 M{\"u}nster,
Germany; Email:~wilka@uni-muenster.de; Phone: +49 251 8334974; 
Fax: +49 251 8334962.}

\begin{abstract}
We study the electron/pion identification performance of the ALICE Transition~Radiation~Detector~(TRD) prototypes using a~neural~network~(NN)~algorithm. Measurements were carried out for particle momenta from 2 to 6\,GeV/$c$. An improvement in pion rejection by about a factor of 3 is obtained with NN compared to standard likelihood methods.
\end{abstract}


\end{frontmatter}

\section{Introduction} \label{d:intro}

The ALICE Transition Radiation Detector (TRD) \cite{aa:tdr} is designed
to provide electron identification and particle tracking in the
high-multiplicity environment produced by heavy-ion collisions at the LHC.
In order to fulfill the envisaged design specifications of the detector, accurate pulse height 
measurement in drift chambers operated with Xe,CO$_2$(15\%) for the duration of the drift time of about 2\,$\mu$s is a necessary requirement.
For electrons, conversions of transition radiation photons~(TR) produced in the 
radiator, are superimposed on the usual ionization energy loss. This is
the crucial factor for improving the electron/pion separation.
A factor of 100 pion rejection for 90\% electron identification efficiency\footnote{Unless otherwise stated, pion rejection factors are quoted or shown for 90\%~electron efficiency.} is the design 
goal of the ALICE TRD consisting of 6 layers. This has been achieved in measurements with prototypes
\cite{aa:id}.

Employing the drift time information in a bidimensional likelihood 
\cite{aa:hol}, the pion rejection capability can be improved by about 60\%
\cite{aa:id} compared to the standard likelihood method on total deposited charge. This method is the simplest way of extending the standard method. However, it does not exploit all recorded
information, namely the amplitude of the signal in each timebin.
Along a single particle track this information is highly correlated due to: 
i) the intrinsic detector signal, in particular since a Xe-based mixture 
is used; 
ii) the response of the front-end electronics used to amplify the
signals.
Under these circumstances, the usage of a neural network (NN) algorithm is a natural choice for the analysis of the data~\cite{aa:hec90,aa:roj93,aa:mul95}. Neural networks are used for a variety of tasks in modern particle detectors \cite{aa:denby}.
A first NN analysis for electron/pion identification with a TRD \cite{aa:bel}
showed that the performance can be significantly improved.

We report results for pion rejection using a NN, which increases the pion rejection factor up to about 500 for a momentum of 2\,GeV/$c$.
The experimental setup and method of data analysis are described in the 
next section. 
We then present the basic ingredients of the analysis and the topology of the neural network we have employed. 
The results are presented in Section~\ref{d:res}.

\section{Experimental Setup} \label{d:meth} 

\begin{figure}[t]
\centering\includegraphics[width=.65\textwidth]{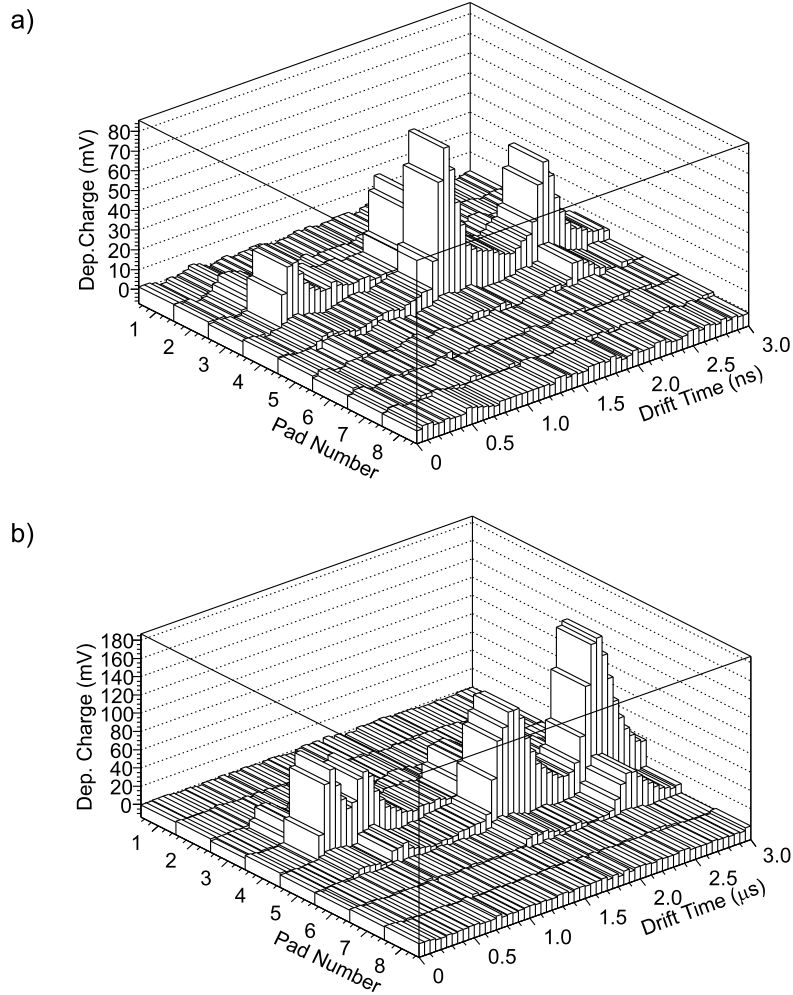}
\caption{The time dependence of TRD signals on eight readout pads a)~for a pion event and b)~for an electron event. The time resolution is 50~ns per time bin, which corresponds to a sampling frequency of 20~MHz. Signals from TR photons are predominantly expected at high time bin numbers.}
\label{d:e10}
\end{figure}

\begin{figure}[t]
\centering\includegraphics[width=.65\textwidth]{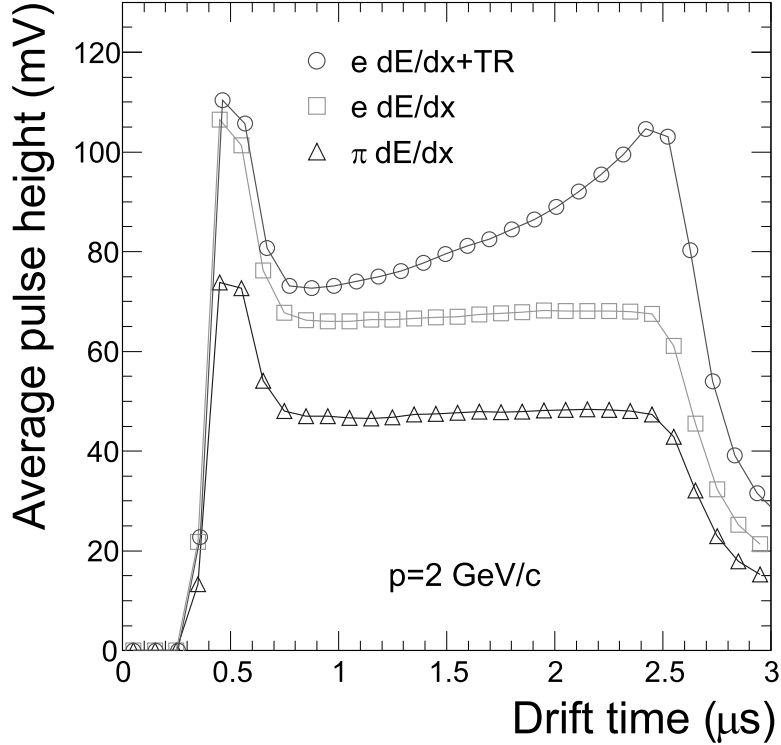}
\caption{Average pulse height for pions and electrons (with and without TR). The average dE/dx~signal for electrons is larger than that for pions. A further increase of the electron signal results from TR photons.}
\label{d:e9}
\end{figure}

For the test measurements, prototype drift chambers (DC) of
the  ALICE TRD~\cite{aa:tdr} were used.
The prototypes have a drift region of 30\,mm and an amplification region 
of 7\,mm.
Signals from induced charge on a segmented cathode plane with rectangular pads 
of 8\,cm length and 0.75\,cm width were recorded.
The drift chambers were operated with the standard gas mixture for the TRD, 
Xe,CO$_2$(15\%), at atmospheric pressure, at a gas gain of about 4000.
For our nominal drift field of 0.7\,kV/cm, the detector signal is spread
over about 2~$\mu$s and is readout using a Flash ADC (FADC) system.
The FADC has an adjustable baseline, an 8-bit non-linear conversion
and 20\,MHz sampling frequency.
The sampling can be rebinned in the off-line analysis to obtain the 100\,ns 
time bins envisaged for the final ALICE TRD design~\cite{aa:tdr}. 
The data acquisition (DAQ) is based on a VME event builder and was developed 
at GSI \cite{aa:mbs}. 

Figure~\ref{d:e10} shows the time dependence of the signal on eight readout pads. Shown is a typical pion event and a typical electron event. For the electron one can see a big cluster at large drift times, possibly produced by TR. The average pulse height versus drift time is shown in Figure~\ref{d:e9}. 
Owing to their larger Lorentz-$\gamma$ electrons deposit more energy in the drift chamber than pions with the same momentum. The peak at small drift~times originates from the amplification region. In a TRD~module the predominant absorption of TR~photons at the beginning of the drift volume leads to an additional peak for electrons at large drift times.

Four identical drift chambers were used in the beam measurements 
with identical radiators in front \cite{aa:id}.
Measurements were carried out at momenta of 2, 3, 4, 5, 
and 6\,GeV/$c$ at the T10 secondary beamline of the CERN PS \cite{aa:cernpi}.
The momentum resolution of this beam was $\Delta p/p\simeq 1\%$.
The beam contained a mixture of electrons and negative pions.
For the present analysis clean samples of pions and electrons were selected
using coincident thresholds on two Cherenkov detectors and a lead-glass 
calorimeter \cite{aa:andr}. Part of the data was taken with a scale down factor for pions, which allowed to get comparable statistics for pions and electrons. Without the scale down factor the number of pions exceeded the number of electrons by a factor of about 15 for a momentum of 6~GeV/$c$.
The incident angle of the beam with respect to the normal to the anode wires 
(drift direction) is 15$^\circ$ to avoid gas gain saturation due to space 
charge \cite{aa:gain}.
For more details on the experimental setup see Ref.~\cite{aa:gain}.

\begin{figure}[t]

\centering\includegraphics[width=.85\textwidth]{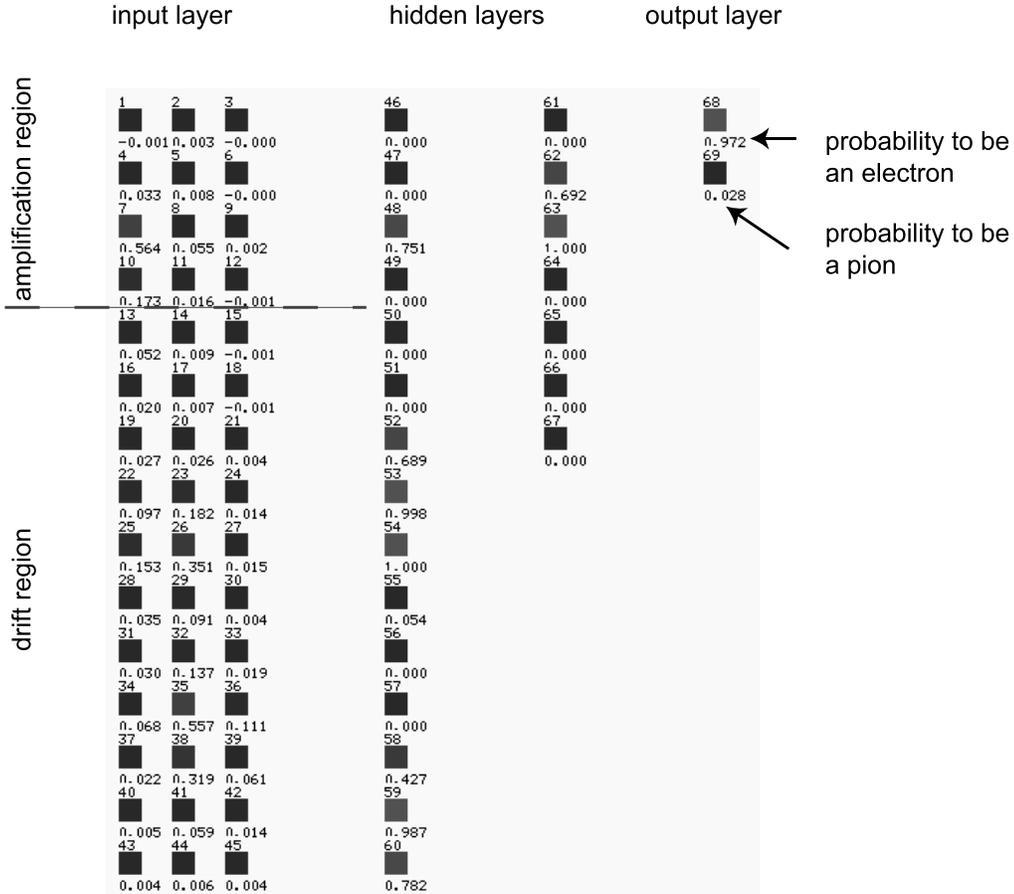}
\caption{NN topology for one TRD chamber. The left side is the input layer, in the middle the hidden layers and on the right side is the output layer. Note, that each neuron of a layer is connected with each neuron in the neighboring layers. For sake of clarity the connections are not drawn. The number below each neuron is the current excitation. Shown is a typical electron event. Three big clusters were generated (at neurons number 7, 26 and 35). The output layer displays the probabilities. In the present event the particle has been classified as an electron with a probability of 97.2~\%.}
\label{d:e1}
\end{figure}

\section{Neural Network Description} \label{d:nn}

The NN analyses were made using the \textit{Stuttgart Neural Network Simulator} (SNNS)~\cite{aa:snns}. This software is equipped with a graphical user interface, which provides several different network types, learning algorithms and analysis tools.  

\subsection{Neural Network Topology for one TRD Chamber} \label{d:nntop}

As described above, electrons and pions produce different patterns in the drift chambers, the NNs should allow efficient pattern recognition. From practical experience it is known that feed forward networks using the back propagation algorithm are particularly suitable for pattern recognition.
We have rebinned the FADC data to 200\,ns per timebin. More than one
pad should be used as input for the neural network, because the signal
spreads over several pads due to charge sharing. In the main analysis
a 15$\times$3 matrix of 15~timebins and 3~pads is used as the input
vector. The pads that were used were the pad with the largest
deposited charge and the two pads on either side of it. Since the usage of raw data would lead to a full activation of all input neurons, the FADC data were normalized to the maximum possible input. This was necessary to guarantee that no input neuron reaches full activation, which would diminish particle separation. Other input topologies are discussed in Section~\ref{d:top}.

Neural networks were trained for single chamber, because it is impractical to generate a network for all chambers. This is due to the general behavior of NNs\footnote{The larger a NN, the longer it takes to train it and the harder it is to find the most generalized network.}. 
A generally accepted indicator for the performance of a neural network is the \textit{mean square error} (MSE). In order to find an adequate topology for the hidden layers, different networks with none, one, two and three hidden layers were trained and tested. Increasing the number of neurons and especially the number of hidden layers generally leads to a smaller MSE. A neural network with three hidden layers reached the smallest value of MSE for training data. The value of the MSE for the test data was comparable to the MSE value of the network with only two hidden layers. In order to prevent \textit{overtraining}, it was decided to use a network topology with two hidden layers. Overtraining leads to a loss in generalization ability and drives the NN to only recognize known patterns.

The final network topology for one chamber is shown in Fig.~\ref{d:e1}. The network is composed of an input layer (45 neurons), two hidden layers (15 and 7 neurons) and an output layer (one neuron for the probability to be an electron and one for it to be a pion). 
Data samples with a pion scale down factor were used to train the NNs. Samples with the same momenta but without pion scale down were used to test the generalization of the NNs and for validation. The learning was done using the \textit{online back~propagation} algorithm. For 3000 epochs and a size of 50,000~events the training took one~hour per chamber with a P4~computer~(2.4~GHz, 256~MB RAM). The validation with 20,000 events took less than a second.

\subsection{Pion Efficiency for Six Chambers} \label{d:PEffi6}

\begin{figure}[t]
\centering\includegraphics[width=1.0\textwidth]{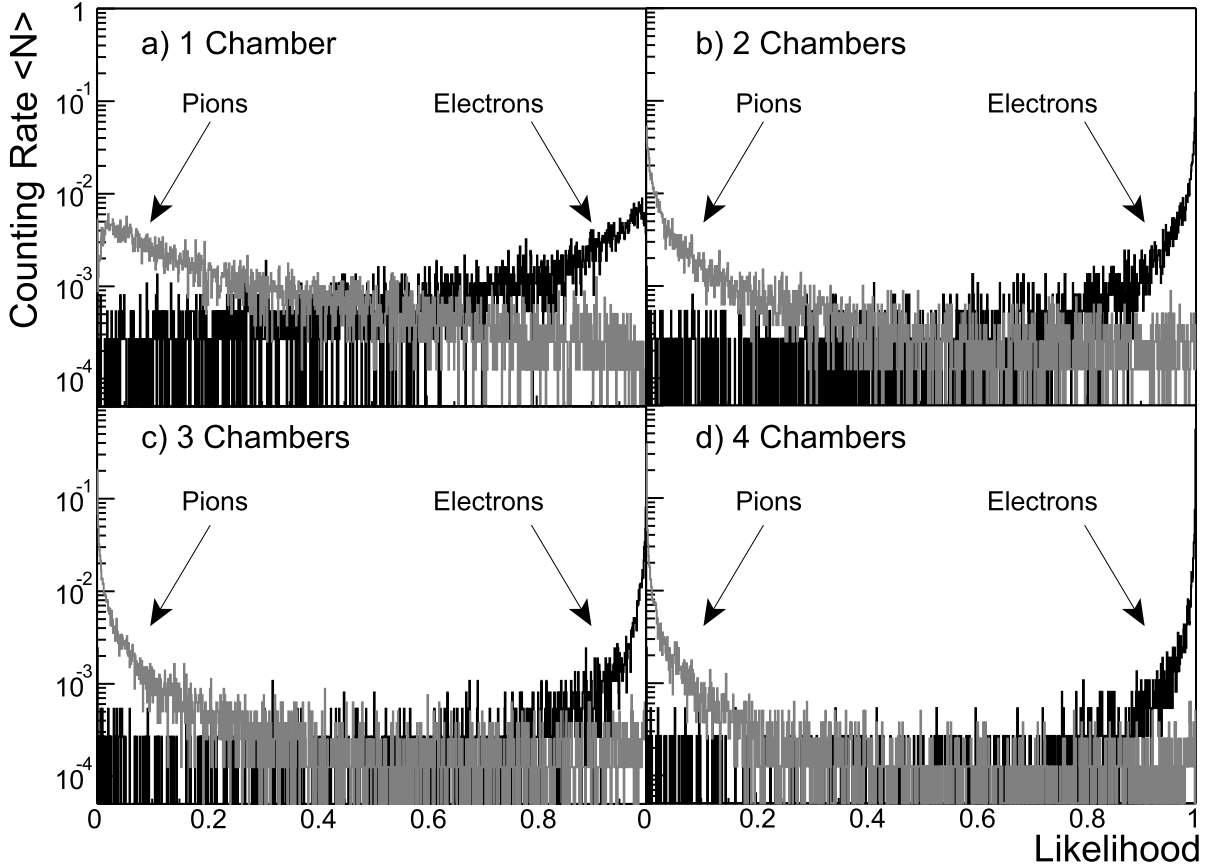}
\caption{Likelihood (to be an electron) distributions for pions and electrons
for the momentum of 2 GeV/c for 1, 2, 3 and 4 chambers.
Electron events are plotted in black and pion events in grey.
}
\label{d:e4}
\end{figure}

The next step was to combine the output of the single chamber networks
to get the pion efficiency for 6 chambers. 
For this we have tested three possibilities: i) a NN with one hidden
layer; ii) a NN without any hidden layer; iii) combining the single chamber
probabilities $P(X_i|e)$ and $P(X_i|\pi)$ into the likelihood (to be an
electron) defined as:
\begin{equation}
\mathrm{L}=\frac{P_e}{P_e+P_\pi}, \quad
P_e=\prod_{i=1}^NP(X_i|e), \quad P_\pi=\prod_{i=1}^NP(X_i|\pi)
\label{TRD:pid_eq1}
\end{equation}
where the products run over the number of detector layers,
$N$. Although the results are very similar in the three cases, the likelihood
method (case iii) gives slightly better pion rejection and, as it is
also the simplest method, will be employed for the rest of the
paper. For each event the probabilities for 1, 2, 3 and 4 chambers
were calculated. The resulting likelihood distributions are shown in
Fig.~\ref{d:e4}. 

\begin{figure}[t]

\centering\includegraphics[width=.75\textwidth]{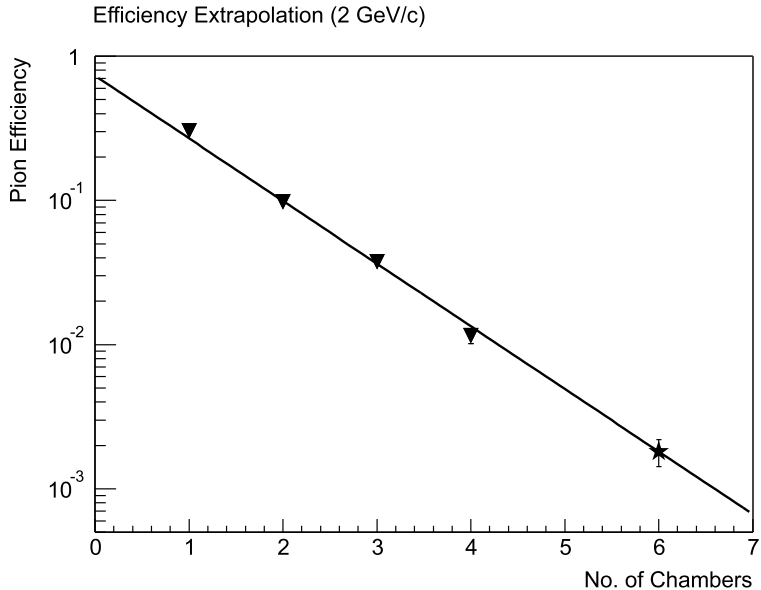}
\caption{Pion efficiency as a function of the number of chambers included
in the likelihood calculation for the momentum of 2 GeV/c. The curve is
a fit to the measured values (triangles). The star shows the pion
efficiency extrapolated for six layers.}
\label{d:e2}
\end{figure}

The pion efficiencies for 1, 2, 3, and 4 chambers are shown in
Figure~\ref{d:e2}. Since the measurements of individual chambers are
uncorrelated one can perform an exponential fit to estimate the pion
efficiency of 6 chambers (star). This corresponds to the expected
performance of the ALICE TRD. The value of the pion efficiency
extrapolated in this way will be used in the following plots.

\subsection{Error Determination} \label{d:ErrBar}

For the described analysis procedure with neural networks it is not straightforward to assign errors to the measured pion efficiencies. A clean, albeit impractical, procedure would be subdivide the measured events into several samples, calculate the pion efficiencies for each sample and finally exploit the variation of the results for the error calculation. As the data size was not large enough for such a procedure, the following approach was used to estimate the uncertainty. 

The data samples were divided into two sets. Pion efficiencies were calculated and compared to the efficiencies obtained with the whole data sample. The maximum difference was used as the error for the pion efficiencies. 

\begin{figure}[t]
\centering\includegraphics[width=.75\textwidth]{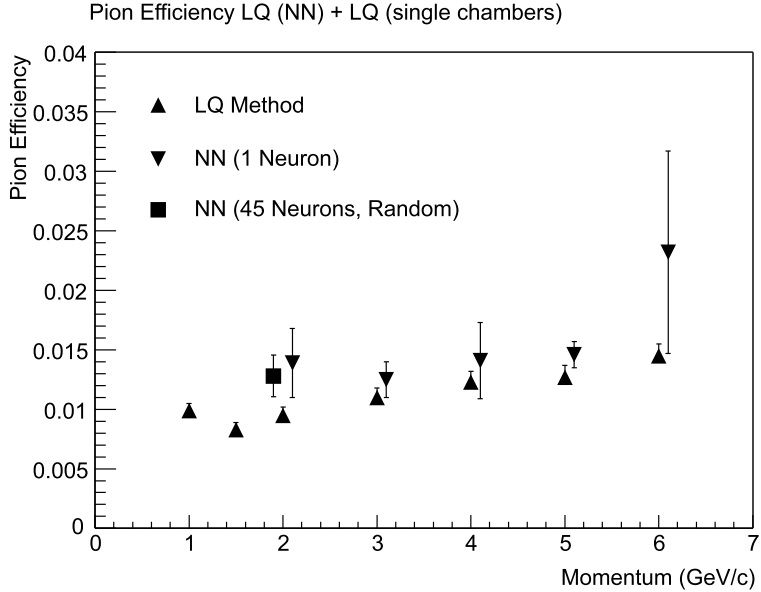}
\caption{Momentum dependence of the pion efficiency. Comparison of the results achieved with the LQ method, with a NN algorithm (with a single input neuron) and the "NN-Random" algorithm.}
\label{d:e5}
\end{figure}

\section{Results and Discussion} \label{d:res}

\subsection{Time-Integrated Signal Analyzed with a~NN and the Classical Likelihood~Methods} \label{d:LQ-NN}

As a first step it was tested whether results obtained with classical methods could be reproduced using a NN. The same network topology as in the analysis described above was used (for the hidden layers and output layer) to guarantee identical conditions. 

Like in the classical \textit{one}-dimensional likelihood method
(\textit{LQ method}\footnote{For further details about the classical
  likelihood methods see \cite{aa:andr}, \cite{aa:zeus},
  \cite{aa:bun}.}), which is based on the total deposited charge, the
charge signal was integrated over all time bins and on three adjacent pads. Thus, the neural network for each chamber has only one input neuron. The learning parameter $\eta$ (which is a factor that describes the velocity of the learning process)\footnote{See \cite{aa:hec90}, \cite{aa:roj93}, \cite{aa:mul95}.} was 0.001 and the networks were trained for 1000 epochs. As expected the results of this method are in good agreement with the results obtained with the LQ method (see Fig. \ref{d:e5}).

A second test, also based on a NN which only considers the integrated deposited charge, was carried out for a momentum of 2~GeV/$c$. Here, the same NN topology like in the main analysis with an identical input layer was used. The position of the different input neurons for each event was randomized, i.e. all information about the position was lost. The NN could obtain the necessary information for its decision whether a particle was an electron or a pion only from the sum of all inputs. The networks were trained for 3000\,epochs\footnote{It was necessary to train the NN for a larger number of epochs compared to previous test, because the input layer was larger and the NN needed more time to adapt.}. The result of this method is, as expected, in good agreement with the LQ and the one neuron results (also Fig. \ref{d:e5}).


\subsection{Pion Rejection with Neural Networks} \label{d:pir}

Here, we present the results of the main analysis, the performance of a neural net algorithm compared to the LQX~method\footnote{Likelihood on total charge and position of the maximum cluster.}. The procedure for this analysis and the used network topology are described in Section~\ref{d:nn}. The networks for the single chamber were trained for 3000 epochs.

\begin{figure}[h]

\centering\includegraphics[width=.75\textwidth]{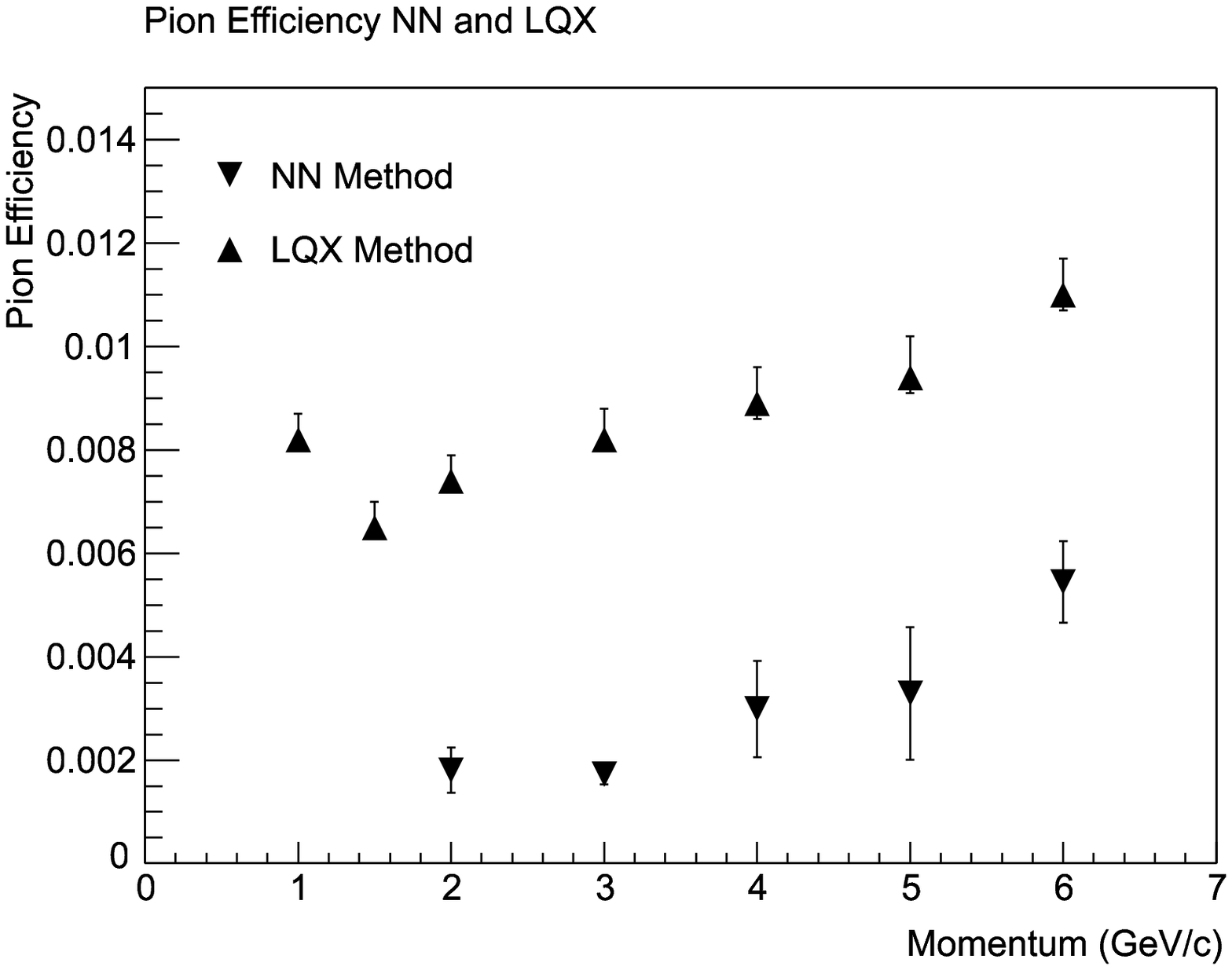}
\caption{Momentum dependence of the pion efficiency for the NN method and for
a bidimentional likelihood (LQX).}
\label{d:e7}
\vspace{2cm}

\centering\includegraphics[width=.75\textwidth]{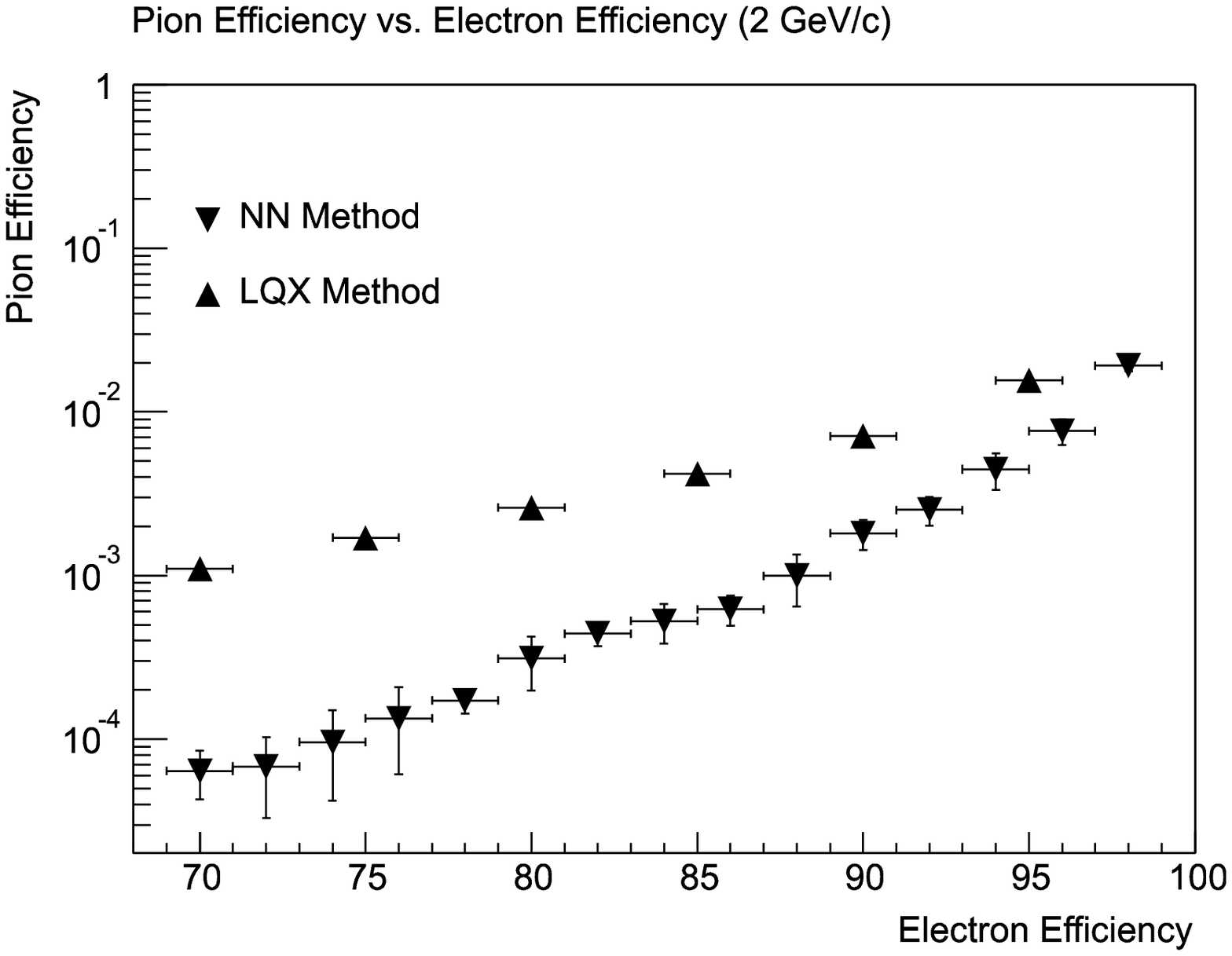}
\caption{Comparison of the results achieved with the LQX method and with a NN algorithm.}

\label{d:e6}
\end{figure}

In Figure~\ref{d:e7} the results of this analysis are presented and
compared to the results obtained with the LQX method. The Figure shows
that using a NN algorithm the pion efficiency is significantly smaller
compared to the LQX method leading to an improvement in the electron/pion separation. For small momenta the improvement nearly reaches a factor of four. As expected (and shown for the LQX method in previous analyses, e.g.~\cite{aa:id}) the pion efficiency increases with higher momenta,thus decreasing the pion rejection factor. This is expected, given that with increasing momenta pions deposit more charge inside the drift chambers which makes it more difficult to separate them from electrons. 

In Figure~\ref{d:e6} the evolution of the pion efficiency for different electron efficiencies for a momentum of 2\,GeV/$c$ is shown. The difference between NN and LQX increases with decreasing electron efficiency.

\subsection{Other Network Topologies} \label{d:top}

There is a large difference between the results of NNs with only one input neuron (section \ref{d:LQ-NN}) on the one hand, and the results of NNs used in the main analysis (section \ref{d:pir}) on the other hand. Hence, the question arises, how the evolution of the pion efficiency from a value of about 1.4~\% to a value of about 0.2~\% for a momentum of 2\,GeV/$c$ depends on the time bin resolution presented to the NN. Depending on the chosen resolution the number of time bins varies. Therefore, the topologies for the input layers differ as well. The topologies of the hidden layers and the output layer are kept unchanged. The number of time bins was varied from 1 time bin (3~$\mu$s width, the NN with one input neuron) to 30 time bins (100\,ns width, the maximum time resolution of the ALICE TRD). The different NNs were trained for between 1000 and 3000\,epochs, depending on the number of time bins. 

\clearpage

\begin{figure}[t]
\centering\includegraphics[width=.75\textwidth]{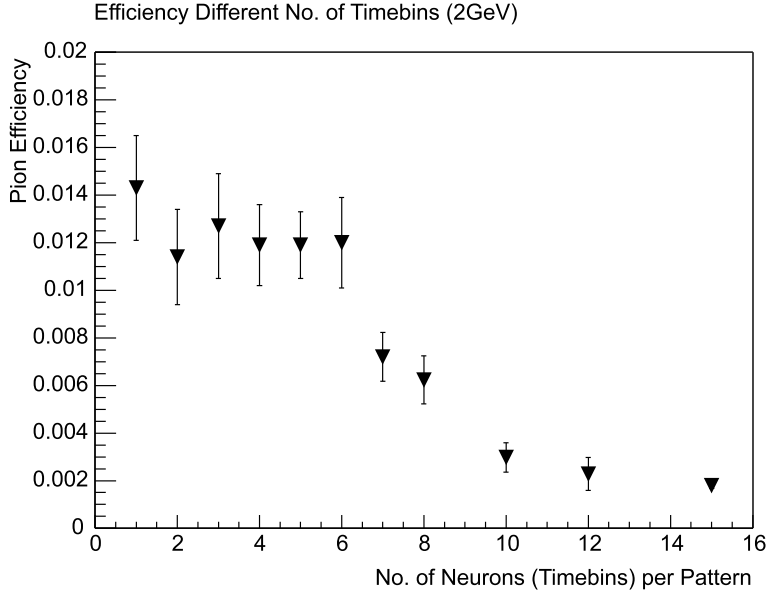}
\caption{Pion efficiencies for different topologies of the input layer for a momentum of 2\,GeV/$c$. On the x-axis the number of input neurons (= timebins) is plotted.}
\label{d:e3}
\end{figure}

Figure~\ref{d:e3} summarizes the results of these analyses. It was expected that the more input neurons were used, the better the performance of the NN algorithm gets. 

This is observed in Fig. 9, however, the pion efficiency is roughly
constant for small number of neurons. After a decrease between 6 and
10~input neurons, the pion efficiency saturates for a larger number of
input neurons as expected.
The result for 30~input neurons is not shown in the graph. It agrees well with the results for 10, 12 and 15 input neurons. 

\section{Summary} \label{aa:sum}

We reported the electron/pion identification performance using data measured with prototype drift chambers 
for the ALICE TRD.
Pions and electrons with momenta from 2 to 6\,GeV/$c$ were studied to analyse the possibilities of a neural network algorithm for electron and pion separation. It was shown that NNs improve the pion rejection significantly by a factor larger than 3 for a momentum of 2\,GeV/$c$ compared to other methods. The rejection power was studied with respect to the number of input neurons and the robustness of the method for electron/pion separation for particles with different momenta was examined. The application of the NN method for simulated full physics events in ALICE TRD is under investigation.

\end{document}